%% file: main.tex
\documentclass[sigconf]{acmart}

\acmConference[Auditing AI Investment Recommendations as Executable Actions]{Anonymous Review}{2026}{ActionAudit}

\acmDOI{}
\acmISBN{}
\pdfmapfile{+libertine.map}
\microtypesetup{expansion=false}
\renewcommand\footnotetextcopyrightpermission[1]{}
\usepackage{pgfplots}
\pgfplotsset{compat=1.18}
\usetikzlibrary{arrows.meta,positioning,fit,backgrounds,calc}
\definecolor{arenaGreen}{HTML}{3D7157}
\definecolor{arenaBlue}{HTML}{466A8D}
\definecolor{arenaRed}{HTML}{A14C5D}
\definecolor{arenaGray}{HTML}{667085}
\definecolor{arenaLight}{HTML}{F5F7FA}
\definecolor{arenaInk}{HTML}{344054}

\title{Auditing AI Investment Recommendations as Executable Actions}

\author{Sidnei Barbieri}
\affiliation{%
  \institution{Aeronautics Institute of Technology}
  \city{São José dos Campos}
  \state{SP}
  \country{Brazil}}
\email{sidneisb@ita.br}

\author{Wellington Vargas}
\affiliation{%
  \institution{Portfel Financial Wealth}
  \city{Barueri}
  \state{SP}
  \country{Brazil}}
\email{wellington.vargas@portfel.com.br}

\author{Ágney Lopes Roth Ferraz}
\affiliation{%
  \institution{Aeronautics Institute of Technology}
  \city{São José dos Campos}
  \state{SP}
  \country{Brazil}}
\email{roth@ita.br}

\begin{abstract}
AI systems increasingly produce investment recommendations, yet the usual evaluations ask the wrong question. Realized return is noisy and easy to overfit, and agreement with a reference portfolio can reward advice that cannot be executed. We argue that an AI-generated recommendation should first be audited as an executable financial action, and only then judged on return. We make this concrete with a deterministic, replayable baseline and a protocol that scores any advisor on three properties a single number conflates: validity under portfolio and fee constraints, stability across repeated runs, and agreement with the baseline.

These properties separate cleanly, and agreement is the most misleading in isolation: across a 120-scenario bank, the control that agrees most with the baseline ($0.94$) is admissible in only $0.58$ of its runs, so agreement certifies an invalid action in $42\%$ of them. On an adversarial set, two frontier models are admissible in barely half of their bare-prompt runs and fail on order arithmetic, not judgment; supplying the fee arithmetic deterministically lifts both to near-perfect validity. We make no alpha claim: the baseline is a transparent verifier whose guardrails follow from the fee schedule and whose decisions replay from frozen inputs, and every figure and table regenerates offline from the artifact.
\end{abstract}

\ccsdesc[500]{Computing methodologies~Artificial intelligence}
\ccsdesc[300]{Information systems~Decision support systems}
\ccsdesc[300]{Applied computing~Economics}

\keywords{AI in finance, investment recommendations, benchmark, trustworthy AI, reproducibility, transaction costs}

\begin{document}
\maketitle

\section{Introduction}

Large language models (LLMs) now produce investment recommendations, and recent work evaluates them using realized returns or agreement with a reference portfolio \cite{oh2025alpha, spadea2025flarko}. Both yardsticks are unsafe. Return is noisy and invites overfitting; agreement rewards an advisor for matching a reference even when its recommendation is operationally unusable or unstable across runs. Parallel evidence shows learned advisors are inconsistent under irrelevant changes and carry latent bias \cite{chawla2025riskadvice, lee2025bias}, which a return or overlap score cannot detect.

We take a different stance: evaluate an advisor against a \emph{deterministic, replayable baseline} on three axes that a single number conflates: operational \emph{validity} under explicit constraints, \emph{stability} across repeated runs, and \emph{agreement} with the baseline. The baseline is a transparent quality, moat, and compounding policy whose recommendations are pure functions of their inputs, so any run can be replayed and diffed. Its trustworthiness as a yardstick rests on reproducibility, not on a claim to beat the market. More generally, we treat an AI advisor as a candidate generator whose output must satisfy an operational contract before its return means anything, and we use investment advice as the domain because that contract is explicit and machine-checkable.

%The paper makes three contributions. First, it adapts behavioral testing to investment advice: an audit protocol that measures validity, stability, and agreement separately, and shows that agreement with a reference can certify an operationally invalid action, in deterministic controls and at scale in a 120-scenario benchmark. Second, it provides the deterministic baseline that makes the audit replayable\footnote{Anonymous reproducibility artifact: \url{https://anonymous.4open.science/r/deterministic-replay-audit/}. Released publicly after the double-blind review.}: the fee model is subadditive, the minimum order follows directly from the cost tolerance, and the planner is regression-tested against those facts. Third, it calibrates the baseline against allocator baselines instead of treating the policy as a performance claim. The current basket beats the market benchmark, but equal weighting is stronger, and risk parity nearly ties it on Sharpe after removing look-ahead from the allocator baselines. This result sharpens the paper's claim: the contribution is verification of investment advice, not a new stock-selection factor.

The paper makes three contributions. First, it adapts behavioral testing to investment advice: an audit protocol that measures validity, stability, and agreement separately, and shows that agreement with a reference can certify an operationally invalid action, in deterministic controls and at scale in a 120-scenario benchmark. Second, it provides the deterministic baseline that makes the audit replayable\footnote{Reproducibility artifact: \url{https://github.com/sidneibarbieri/deterministic-replay-audit}}: the fee model is subadditive, the minimum order follows directly from the cost tolerance, and the planner is regression-tested against those facts. Third, it calibrates the baseline against allocator baselines instead of treating the policy as a performance claim. The current basket beats the market benchmark, but equal weighting is stronger, and risk parity nearly ties it on Sharpe after removing look-ahead from the allocator baselines. This result sharpens the paper's claim: the contribution is verification of investment advice, not a new stock-selection factor.

\section{Related Work}
\label{sec:related}

Our work sits at the intersection of four lines. Factor research establishes durable signals for security selection: gross profitability \cite{novyMarx2013}, the quality-minus-junk composite \cite{asness2019qmj}, and high-dimensional machine-learning predictors in the cross-section \cite{gu2020machine}; relatedly, firms with a durable competitive advantage, a wide economic moat, earn stronger long-run risk-adjusted returns \cite{kanuri2016moat}, which is why the baseline scores moat alongside quality and compounding. We treat selection as given and do not propose a new factor; our scoring reuses such established signals only to rank an existing basket. The choice of signals is not load-bearing: the audit accepts any deterministic policy as its yardstick, and we report that a parameter-free equal-weight allocator already matches ours; thus, substituting momentum, profitability, or low volatility would not alter the protocol. Our claim is orthogonal to stock-picking: it is the layer that audits \emph{whether advice can be executed} in a replayable and cost-effective manner once a candidate set exists.

Classical theory shows that transaction costs induce a \emph{no-trade region}: proportional costs make the optimal policy a band around the target where one does not rebalance \cite{constantinides1986capital, davis1990portfolio}. With multiple assets, costs produce explicit no-trade geometry \cite{liu2004transaction}, partial adjustment under predictable returns \cite{garleanu2013dynamic}, mixed-integer or convex programs \cite{lobo2007portfolio}, and liquidity or market-impact terms \cite{zhang2019dynamic}. Fixed-fee rebalancing can also be formulated as a linear programming problem or a heuristic search \cite{delarosa2023planning}. Our objective is narrower and more auditable: recurring small-cash deployment under a quantized step fee. The fixed-cost guardrail is the small-cash analog of the no-trade region: below the floor, the optimal action is to hold.

A fast-growing line uses language models to recommend assets or build portfolios and evaluates them almost entirely by realized return or alignment with a reference: media-driven LLM portfolios report strong but short-horizon gains \cite{oh2025alpha}, and knowledge-graph-aligned recommenders are scored on behavioral alignment \cite{spadea2025flarko}. That verdict depends sharply on the evaluation window. A live, two-month, multi-market arena finds LLM agents can beat buy-and-hold by exploiting short volatility bursts rather than long trends \cite{qian2026ama}. A contamination-free, multi-month benchmark tempers this, since most agents fail to beat buy-and-hold, and strong financial question-answering does not transfer into trading skill \cite{chen2025stockbench}. A two-decade, hundred-symbol evaluation reverses it outright once survivorship, look-ahead, and data-snooping biases are removed \cite{li2026finsaber}. That long-horizon study is also diagnostic. The agents are \emph{miscalibrated} rather than merely weak, too conservative in bull markets and too aggressive in bear markets, a plain ARIMA predictor dominates them, and the binding constraint is a missing layer of domain-aware financial logic, not model scale. Two signals survive across these otherwise conflicting studies. First, both the arena and the long-horizon study attribute behavior chiefly to the agent framework rather than the model backbone \cite{qian2026ama, li2026finsaber}, so the object worth auditing is the advice an agent emits, not the model alone. Second, because the return verdict flips with horizon and regime, no single submission can settle an advisor on return. We take the complementary path: evaluate advice where it is cheap, exact, and replayable, against a deterministic baseline that supplies the domain-aware financial logic the agents lack, before any return is computed; the backtest inherits the same discipline, estimating its allocator baselines walk-forward from past-only windows so no future information leaks in \cite{bailey2014pseudo}. Field evidence on robo-advice is consistent: automated advice can dampen behavioral biases without eliminating them \cite{dacunto2019robo}.

A complementary line documents failure modes that a return or overlap score cannot see. Finance trustworthiness benchmarks now assess truthfulness, safety, fairness, privacy, and transparency \cite{hu2025fintrust}, yet they grade single-turn responses rather than the admissibility of the resulting trade. Standardized financial-LLM benchmarks reinforce the point: PIXIU with its FLARE suite \cite{xie2023pixiu} and the broader FinBen \cite{xie2024finben} grade dozens of tasks from sentiment to stock-movement and even trading, yet score accuracy on text, not whether a recommended action is admissible and executable. The gap is operational. LLM risk assessments shift in response to irrelevant demographic attributes \cite{chawla2025riskadvice}, investment analyses exhibit latent, confirmation-seeking bias \cite{lee2025bias}, and recommenders show product preferences that persist after debiasing \cite{zhi2025productbias}, while surveys flag interpretability and generalizability as open problems \cite{elalami2025mlfinance}. This stance has a lineage outside finance: holistic evaluation replaced a single accuracy number with parallel desiderata \cite{liang2023helm}, behavioral testing replaced aggregate metrics with capability-specific unit tests \cite{ribeiro2020checklist}, and agent benchmarks moved the unit of evaluation from a correct answer to a correct action \cite{liu2024agentbench, jimenez2024swebench}. In contrast, financial-LLM evaluation has not made this move. Our protocol is behavioral testing specialized to a financial action: each validity check is a capability test, and the failure taxonomy is its test matrix, run deterministically against a frozen scenario. The ungrounded-fact check is the verifiable end of faithfulness evaluation, which hallucination research still treats as largely open \cite{ji2023hallucination}. What no prior benchmark supplies is a compact, replayable test that places a learned advisor against a transparent baseline and measures, together, whether its advice is \emph{valid}, \emph{stable} across runs, and how far it \emph{agrees}; the advisor under test may be an LLM, a robo-allocator, or a heuristic.

Most directly, our verifier belongs to an emerging line on \emph{runtime governance of LLM-generated actions} that intercepts an action before execution and checks it against an explicit policy. AgentSpec enforces a customizable rule language at a pre-execution hook, with a financial transfer as its motivating example \cite{wang2025agentspec}. VeriGuard synthesizes and formally verifies a policy offline, then validates each online action against it \cite{miculicich2025veriguard}. ProbGuard predicts violations ahead of time with probabilistic guarantees \cite{wang2026probguard}. TRAC monitors temporally extended compliance with temporal logic, and reports that small or deterministic labelers match or exceed frontier-model judges \cite{alamdari2026formalmethods}. Our contribution is the exact, self-contained end of this line: a hand-specified admissibility contract for a financial action, referenced to declared quantities, so the verdict is replayable and judgment-free rather than model-synthesized or probabilistic. To these enforcement mechanisms, we add what they presuppose but do not measure: an empirical demonstration, on real frontier models, that agreement and admissibility come apart, and that the dominant source of admissibility failure is computation, not judgment. The same deterministic-labeler result independently supports our choice of a rule engine over a model judge.

A large fraction of reported strategy performance does not survive out of sample; backtest overfitting is easy to produce and hard to detect \cite{bailey2014pseudo}, and naive $1/N$ diversification is a stubborn baseline \cite{demiguel2009naive}. The artifact follows the discipline these results require: the engine is deterministic, every empirical figure is regenerated from it, and the calibration is reported even when simple allocators outperform the tuned weighting.

\section{Artifact and Audit Protocol}
\label{sec:audit}

\begin{figure*}[t]
  \centering
  \resizebox{0.86\textwidth}{!}{\input{figures/audit_boundary.tikz}}
\caption{The audit boundary turns advisor text into executable orders and typed violations before any return is computed.}
\Description{A mechanism diagram showing frozen scenario data and advisor output entering a normalizer and deterministic verifier, which applies policy rules and emits admissible orders, violation traces, and audit metrics.}
  \label{fig:workflow}
\end{figure*}
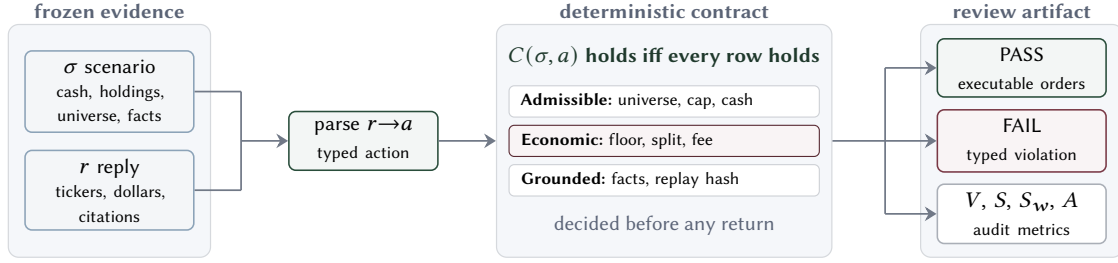

The backend separates API data-transfer objects, portfolio importers, provider adapters, analytics scoring, and deployment planning. Network access is isolated behind provider abstractions. The scoring and deployment modules perform no I/O: a recommendation is a pure function of the tuple.
\[
(\text{holdings},\ \text{fundamentals},\ \text{cash},\ \text{policy version}).
\]
Referential transparency is what makes the rest of this paper possible. Every quantitative result is backed by an offline driver in the reviewer path: \texttt{run\_experiments.py} for the experiment JSON and figures, \texttt{run\_ai\_advisor\_audit.py} for the advisor-audit panel, and \texttt{collect\_advisor\_runs.py} for the frontier-model audit. None depends on live data or hidden state. Purity is enforced, not asserted: a replay test feeds the same input tuple twice and checks that orders, exclusions, and ranked candidates are byte-identical, so any past recommendation can be regenerated and diffed from its logged inputs. Determinism is engineered: the engine draws no random numbers, and ties are broken by ticker. The verdicts avoid fragile floating-point equality, since fees and the floor are integer tranche counts, validity and agreement are set memberships, and stability is a ratio of those sets, so a low-order bit difference cannot flip an outcome. Figure~\ref{fig:workflow} shows where this boundary sits: the model proposes text, while deterministic code emits executable orders or typed violations.

The composite score is the bounded linear functional $s=w_m\,\mathrm{moat}+w_c\,\mathrm{comp}+w_v\,\mathrm{val}$, with sub-scores in $[0,100]$ and weights $(w_m,w_c,w_v)=(0.40,0.35,0.25)$. Holdings already above $1.3\times$ equal weight, or in a theme at its concentration cap, are excluded; the planner then ranks the remainder by $s$ and sizes orders against available cash.

We evaluate an advisor not by how well it agrees with the baseline but by three axes a single number conflates. Fix a \emph{scenario}: current holdings, available cash, an allowed universe, per-scenario constraints (a recommendation cap, add-only or trim rules, the economic order floor of Section~\ref{sec:guardrail}), and the baseline's own recommendation. An advisor returns a set of tickers (optionally with dollar amounts and cited facts). We score three quantities. \emph{Validity} is the fraction of runs with no constraint violation: no out-of-universe ticker, no cap breach, no add-only ownership violation, no cash overrun, no below-floor order, no fee-worsening split, and no cited fact absent from the frozen scenario. \emph{Stability} is the mean pairwise overlap of repeated recommendations, with an amount-aware sizing score when dollar allocations are returned. \emph{Agreement} is the overlap with the baseline's top-$k$ set, where $k$ is the number of positions the baseline actually funds under the scenario's cash and recommendation cap, not a free parameter. Agreement is reported, not rewarded by itself, because disagreement can be either useful novelty or error. Because it is descriptive rather than optimized, the independence result does not hinge on $k$: validity and stability are intrinsic to the advisor and unchanged by the choice, so a different $k$, or a rank-correlation variant, would rescale agreement without altering which axis a failure appears on.

The validity checks form a deterministic taxonomy of failures (Table~\ref{tab:taxonomy}). Each category is an exact test on the advisor's output and the frozen scenario, so a failed run is attributable to a named cause with no second model in the loop, which is what keeps the verdict replayable and free of judgment. The categories are not hypothetical: the sub-tranche control triggers the fee-worsening split, the portfolio-fit case triggers the concentration breach, and the fact-hallucination control triggers the ungrounded fact.

\begin{table}[tb]
\caption{Deterministic failure taxonomy: each category is an exact test on advisor output and the frozen scenario ($a_i$ is leg $i$'s amount, $T$ the fee tranche, $c/\tau$ the economic floor).}
  \label{tab:taxonomy}
\begin{tabular}{ll}
\toprule
Failure & Deterministic test \\
\midrule
Out-of-universe ticker & ticker $\notin$ allowed set \\
Recommendation-cap breach & count $>$ cap \\
Already-owned (add-only) & ticker $\in$ held set \\
Cash overrun & $\textstyle\sum_i a_i >$ cash \\
Below-floor order & $a_i < c/\tau$ \\
Fee-worsening split & $\textstyle\sum_i \lceil a_i/T\rceil > \lceil (\textstyle\sum_i a_i)/T\rceil$ \\
Concentration breach & post-trade weight $>$ cap \\
Ungrounded cited fact & fact id $\notin$ scenario set \\
\bottomrule
  \end{tabular}
\end{table}

A single trace fixes the mechanism. In the sub-tranche scenario the advisor has \$900 to deploy under a \$2.50 fee per \$1{,}000 tranche. One \$900 order is admissible and pays a single tranche fee; a bare model instead splits the cash into two \$450 legs. The verifier evaluates the split row of Table~\ref{tab:taxonomy}, $\lceil 450/1000\rceil + \lceil 450/1000\rceil = 2 > 1 = \lceil 900/1000\rceil$, and returns a fee-worsening split before any price is observed. The advisor's text has become a typed action, the contract has decided it exactly, and the verdict replays from the logged inputs.

\begin{figure}[tb]
  \centering
  \includegraphics[width=\columnwidth]{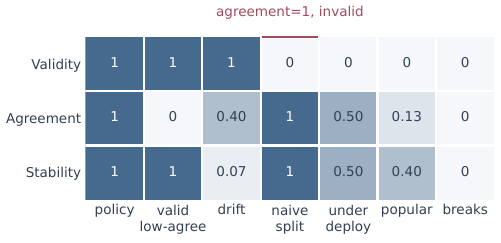}
  \caption{The archetype controls separate validity, agreement, and stability.}
\Description{Heat-map matrix of validity, agreement, and stability across seven advisor archetypes, showing that high agreement does not coincide with high validity.}
  \label{fig:audit}
\end{figure}

The three axes do not substitute for one another. We show this with seven deterministic archetype controls: failure modes by construction, not model outputs (Figure~\ref{fig:audit}). The decisive control is the naive diversifier: it agrees with the baseline and is perfectly stable, yet is invalid because it splits sub-tranche cash and pays an avoidable fee. The mirror image is a valid-but-disagreeing control: fully valid and stable, but with zero agreement. An overlap-only benchmark would rank the first above the second, which reverses their operational quality.

For an advisor that returns ticker sets $r_1,\dots,r_n$ over repeated runs of a scenario with baseline top-$k$ set $b$, the metrics are:
\[
  \begin{array}{r@{\;=\;}l@{\qquad}r@{\;=\;}l}
V & \tfrac1n\sum_j \mathbf{1}[\mathrm{viol}(r_j)=\varnothing] & S & \dbinom{n}{2}^{-1}\!\sum_{i<j} J(r_i,r_j) \\[1.8ex] S_w & 1 - \dbinom{n}{2}^{-1}\!\sum_{i<j} \mathrm{TV}(w_i,w_j) & A & \tfrac1n\sum_j \dfrac{|r_j\cap b|}{k}
  \end{array}
\]

where $J$ is the Jaccard overlap and $\mathrm{viol}$ the constraint set of $r_j$, $w_j$ normalizes run $j$'s dollar amounts to weights, and $\mathrm{TV}$ is total-variation distance. The sizing stability $S_w\in[0,1]$ is reported when a scenario requires amounts. The audit is therefore replayable once the advisor outputs are frozen. It certifies process quality, not future return: advice must be admissible, repeatable, and grounded before performance statistics are meaningful.

The sizing axis earns its place on the controls. In the sub-tranche scenario, the naive diversifier repeats the same tickers on every run, so its set stability is a perfect $1.0$; it nonetheless shuffles the dollar split enough to drop $S_w$ to $0.93$, and an under-deploying control falls to $S_w=0.33$, while the deterministic policy holds both terms at $1.0$. Set overlap alone would certify the diversifier as perfectly stable, so the amount-aware term is what exposes the sizing drift that set overlap cannot see.

The same audit scores the real model output once the responses are frozen. We audit two current frontier models, GPT-5.5 (OpenAI) and Opus-4.8 (Anthropic), under one identical configuration (Table~\ref{tab:fairness}): single-message JSON-only calls, temperature $1.0$, no tools, and no system prompt, so the comparison reflects the models, not the harness. Each record stores the provider, model label, prompt, prompt hash, raw response, parsed output, token usage, finish reason, latency, and timestamp, and is in the artifact's data-hash manifest. One rule governs the experiment: no model output becomes a claim until the harness itself is audited. A tight $500$-token budget initially truncated the reasoning model's completion. The answer was cut off mid-generation and read as an abstention, so we raised the budget until no completion was clipped, and now record a per-run truncation flag. Every reported run satisfies $\textit{truncated}=0$.

Across the three pilot scenarios, both models satisfy all constraints and return a single \$900 order in the sub-tranche case rather than splitting. That pilot is easy by construction, so we also sweep twenty-four adversarial scenarios across three prompt arms, three runs each. Their cash amounts straddle fee tranches, and their fact sets never state the fee, which forces the model to compute the arithmetic rather than recite it. Under a bare prompt, both models are valid in only about half their runs (GPT-5.5 $0.60$, Opus-4.8 $0.53$), and stating the rules as prose barely moves them ($0.58$ and $0.57$). The two failure modes differ but are equally invisible to a return or overlap score. The GPT-class model's invalid runs are all arithmetic: under the bare prompt, $29$ fee-worsening splits and $18$ below-floor orders. Opus-4.8 instead abstains, returning no amounts on $30$ of its bare-prompt runs. The scaffold arm supplies the fixed arithmetic that has already been computed. It removes the GPT-class violations completely and leaves Opus-4.8 a single already-owned slip, lifting validity to $1.00$ and $0.99$. The reading is causal: computation, not judgment, dominates admissibility failure. Prose rules move validity by at most one run, but externalizing the arithmetic makes it near-perfect for both models, so the deterministic layer should own the arithmetic and leave the verifier to guard the lone semantic residual.

\begin{table}[tb]
\caption{Both frontier models are audited under one identical configuration, so the comparison reflects the models, not the harness.}
  \label{tab:fairness}
\begin{tabular}{ll}
\toprule
Parameter & Value (both models) \\
\midrule
Temperature & $1.0$ (provider default) \\
Completion budget & $4000$ tokens \\
Top-$p$ & provider default \\
Tools / system prompt & none \\
Runs per scenario & $3$ \\
Truncated completions & $0$ \\
\bottomrule
  \end{tabular}
\end{table}

The pilot covers three scenarios; the offline scenario bank tests the same separation at scale. The bank contains 120 frozen scenarios, ten variants each of twelve categories spanning cash deployment, ownership, fee sensitivity, grounding, drawdowns, and rebalancing. The fact scenarios use dated market fact identifiers for returns, rates, employment, and concentration claims, so grounding is tested against realistic inputs while remaining a deterministic set check. The oracle is the manifest itself: cash, allowed tickers, baseline tickers, fact IDs, and constraints are frozen before any advisor output is scored. Five deterministic advisor configurations run five times per scenario, for 3,000 replayed outputs. The configurations are synthetic stressors rather than provider claims: one implements the contract, one maximizes agreement while splitting orders, one changes tickers across runs, one cites unsupported facts, and one overspends cash.

\begin{table}[tb]
\caption{Frozen scenario-bank results over 120 scenarios and 3,000 advisor outputs. High agreement and stability do not imply validity.}
  \label{tab:scenario-bank}
\begin{tabular}{lrrrr}
\toprule
Advisor & Valid & Agree & Set $S$ & Agr.-FP \\
\midrule
Contract & $1.00$ & $0.92$ & $1.00$ & $0.00$ \\
Agreement split & $0.58$ & $0.94$ & $1.00$ & $0.42$ \\
Unstable selector & $0.95$ & $0.32$ & $0.13$ & $0.00$ \\
Fact hallucination & $0.67$ & $0.92$ & $1.00$ & $0.33$ \\
Cash overrun & $0.50$ & $0.68$ & $1.00$ & $0.42$ \\
\bottomrule
  \end{tabular}
\end{table}

Table~\ref{tab:scenario-bank} turns the pilot failure into a benchmark result. The agreement-only false-positive rate (Agr.-FP) is the fraction of runs that reproduce the baseline's funded top-$k$ set yet violate a constraint. The agreement-only splitter has higher agreement than the contract and perfect set stability, yet its Agr.-FP is $0.42$, so a top-$k$ benchmark would pass an invalid action in nearly half of its runs. The failure taxonomy is likewise operational: 300 cash overruns, 200 ungrounded facts, 150 fee-worsening splits, 150 too-many-recommendation violations, 100 below-floor orders, and 30 malformed amount outputs. These are deterministic violations of the scenario contract, and the scenario manifest is hashed by the experiment driver. The bank is balanced by construction, ten variants for each of the twelve categories, and because the manifest is hashed, adding or removing a scenario is a versioned change to the benchmark rather than a silent edit.

\section{Deterministic Verification}
\label{sec:portfolio-verification}
\label{sec:fees}
\label{sec:guardrail}

The audit protocol treats an advisor as a candidate generator, not as the final policy. This distinction matters in portfolio advice. A ticker can be attractive in isolation and still be the wrong next action for a specific portfolio because it increases concentration, consumes cash inefficiently, or repeats an exposure that is already full. The deterministic layer, therefore, verifies each candidate against the current portfolio before it can be recommended.

The verifier uses information that the advisor must already have exposed: current holdings, themes, cash, fees, and the candidate score. It estimates the projected theme weight of a starter position, assigns a structural role (diversifier, theme reinforcement, or concentration watch), and ranks the candidate by portfolio fit after the quality score is known. In the controlled scenario used in the artifact, three candidates compete: the highest-quality Platforms name (quality $90$) is flagged a concentration \emph{watch} because it would push an already-crowded theme from $23.0\%$ to $30.7\%$; an Industrial name (quality $84$) is selected as a \emph{diversifier} at a new $10.0\%$ exposure; and a Healthcare name (quality $82$) only \emph{reinforces} an existing $14.5\%$ theme. An isolated rank would buy Platforms; portfolio verification buys Industrial.

This is the core computer-science move in the paper. The learned system may search a large, noisy space of plausible investment theses; the deterministic artifact checks admissibility against a compact contract. The contract is small enough to replay and audit, but strong enough to reject advice that a pure ranking metric would accept.

The contract is explicit, not ad hoc, and every threshold is referenced to a declared quantity. A candidate is admissible only if, after the proposed order, its position weight stays within $1.3\times$ the equal-weight share and its theme stays within a $20\%$ exposure cap; the order is at least the economic floor $c/\tau$; any split is fee-neutral against the consolidated order; and every fact the advisor cites is present in the scenario snapshot. The first two bounds are ratios of equal weight and a single cap, not per-asset tuning, so the same contract applies to any basket. Theme membership is also part of the frozen input, not a live classifier: the artifact uses a versioned ticker-to-theme map and logs the map with the scenario, so changing the taxonomy is a changed experiment. Fact-grounding is likewise a set test, not a language model judgment: the scenario fixes a finite set of citable fact identifiers, and a citation outside that set is a violation, which gives a deterministic check on explanation faithfulness.

The same verifier must translate a ranking into executable orders. That translation matters whenever the broker charges a fixed cost $c$ for each started tranche of size $T$ in a single-symbol order:
\begin{equation}
f(a) = \left\lceil \tfrac{a}{T} \right\rceil c,
  \qquad T = \$1000,\ c = \$2.50.
  \label{eq:fee}
\end{equation}
The ceiling is an important detail. A \$1{,}001 order pays two tranches, the same as a \$2{,}000 order. Since ceilings are subadditive, $f(a+b)\le f(a)+f(b)$: splitting cash across symbols never lowers the fee and can raise it.

We measure the excess fee from splitting as a diversification premium,
\begin{equation}
\pi(a_1,\dots,a_k) = \sum_i f(a_i) - f\!\Big(\sum_i a_i\Big) \;\ge\; 0,
  \label{eq:premium}
\end{equation}
where the right-hand term is the lower bound charged by one order of the same size. This is not a recommendation to buy only one asset; it is a cost reference that indicates when diversification is worth the cost.

\begin{figure}[tb]
  \centering
  \includegraphics[width=\columnwidth]{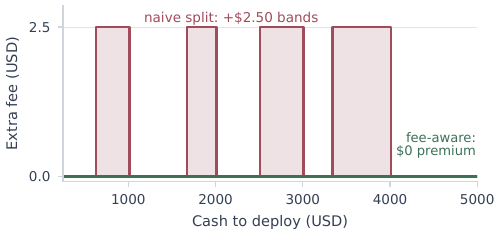}
\caption{Naive splitting pays recurring fee premiums; the fee-aware planner stays on the zero-premium frontier.}
\Description{Line chart showing naive split fee premiums above zero in repeated bands, while the fee-aware planner remains at zero premium.}
  \label{fig:premium}
\end{figure}

We swept cash from \$250 to \$5000 and compared three policies against the single-order lower bound: a \emph{naive} split strictly proportional to score, a \emph{fee-aware} planner, and (as a reference) the single order. Naive splitting incurs $\pi=\$2.50$ in recurring bands wherever the two legs cross an extra tranche boundary (Figure~\ref{fig:premium}).

The sweep exposed and fixed a real planner defect. A score-proportional split can pay an extra \$2.50 when two sub-tranche legs each start a fee unit. The earlier planner made that mistake for cash between \$625 and \$1000 under a 60/40 split. The corrected rule is direct: diversify only when the split is fee-neutral; otherwise, consolidate. A property test now asserts that the planner's fee never exceeds the single-order lower bound across the cash grid.

A second cost rule prevents tiny orders. If a trade should lose at most a fraction $\tau$ to the fixed fee, then the binding first-tranche order is
\begin{equation}
  \mathrm{MIN} = \frac{c}{\tau} = \frac{\$2.50}{0.01} = \$250,
  \label{eq:min}
\end{equation}
Below this floor, the planner holds cash, and a second economic order requires at least \$500. The floor is linear in the fixed fee and inverse in the tolerance, so it transfers across broker fee schedules rather than being a tuned constant. At the same 1\% tolerance, a flat \$1.00 ticket gives a \$100 floor, a \$5.00 ticket a \$500 floor, and a \$9.99 ticket a \$999 floor, while zero commission collapses the guardrail to nothing, which is the correct behavior: when execution is free, there is no reason to wait. The tranche structure changes the fee within an order but not this floor, so a reviewer can stress the schedule and recover the rule directly. The guardrail is the fee schedule made explicit, not a tuned threshold.

The remaining risk is heuristic optimism: a planner can look sensible while leaving deployable cash on the table. We therefore compare it with two simple oracles. The fee oracle is the lower bound on the consolidated-order premium from Equation~\ref{eq:premium}. The deployment oracle is a mixed-integer program (MIP) that selects order sizes and on/off flags to maximize the cash deployed, subject to the same budget, economic floor, theme cap, and per-name headroom constraints as the planner. Concretely, it uses allocation variables $x_i$ and order flags $z_i$, maximizes $\sum_i x_i$, and enforces cash budget, $x_i=0$ unless $z_i=1$, $x_i\ge(c/\tau)z_i$, theme caps, and per-name headroom. The artifact resolves this reference using PuLP/CBC and records the status, time, deployed cash, and order count. It is an upper bound on feasible deployment, not a return forecast.

Across a small-cash grid from \$250 to \$5{,}000 in the planner's design regime, the planner matches both references exactly. It deploys the same cash as the MIP oracle and pays the fee lower bound, so the measured deployment gap and fee premium are both zero. The MIP can deploy more only once a concentration limit is reached and the heuristic stops, which falls outside the recurring small-cash setting this planner targets. For the fee-quantized cash deployments studied here, the planner's cost is established against exact references rather than asserted.

\section{Calibrating the Baseline}
\label{sec:weighting}

The baseline is the audit's yardstick, so this section calibrates the yardstick itself rather than any advisor: it shows that the factor weighting is not load-bearing and that the baseline is a fair, reproducible reference. Because the composite score is linear, the ranking it induces is directly ablatable: zeroing a weight isolates a factor, and the score can be recomposed from per-position sub-scores without re-fetching data.

\begin{figure}[tb]
  \centering
  \includegraphics[width=\columnwidth]{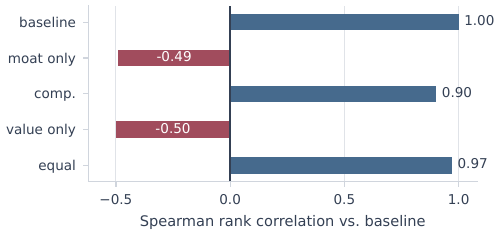}
\caption{Each factor reshapes the ranking differently; equal weighting tracks the tuned baseline.}
\Description{Horizontal bar chart comparing Spearman rank correlation against the baseline across moat-only, compounding-only, valuation-only, equal-thirds, and baseline weighting schemes.}
  \label{fig:ablation}
\end{figure}

On a fixed, deterministic basket, we recomposed the ranking under single-factor and equal weights and measured Spearman correlation to the baseline ordering and the churn in the funded top-2 (Figure~\ref{fig:ablation}). Single factors reorder the basket sharply: moat-only and valuation-only are \emph{anti-correlated} with the baseline, while equal thirds reproduce the baseline almost exactly ($\rho=0.97$, zero churn). The funded picks are stable under $\pm10\%$ weight perturbation. (This basket uses deterministic synthetic fundamentals; it demonstrates the \emph{mechanism}, not a claim about specific securities.)

We backtested the 17-name basket against four baselines over 2021-01-05 to 2026-06-04 (1360 trading days) using free adjusted closes (Table~\ref{tab:backtest}). Two are naive (market-value ``current'' weighting and equal weighting); two are allocator baselines computable from the return matrix alone: long-only minimum variance and equal-risk-contribution (risk parity). To avoid look-ahead, allocator weights are estimated walk-forward from a rolling 252-day return window, use a 20\% constant-variance shrinkage target, rebalance every 63 trading days, and pay 10 bp turnover cost. Until 63 observations exist, the allocator uses equal weights. Minimum variance uses an active-set reduction on the closed-form solution, and risk parity uses the Maillard fixed-point iteration. Reported estimation choices are deliberately plain, since the point is a fair reference, not a tuned competitor.

The score-weighted basket beats the SPY benchmark by $+4.96$ pp in compound annual growth rate (CAGR) and $+0.18$ in Sharpe ratio, but equal weighting is stronger ($22.8\%$ CAGR, $1.19$ Sharpe ratio). The walk-forward risk-parity baseline comes close on Sharpe ($1.11$ versus $1.14$ after rounding) while producing a lower return; minimum variance buys the shallowest drawdown ($-21.7\%$) at a high cost. Periodic rebalancing \emph{reduced} CAGR to $18.8\%$ from the $20.7\%$ buy-and-hold figure despite low explicit cost. The synthetic ablation and the real backtest point the same way: on this basket, selection matters more than a tuned fundamentals weighting scheme, and the paper's contribution is therefore the audit protocol rather than an alpha claim.

\begin{table}[tb]
\caption{Current-basket backtest, 2021-01-05 to 2026-06-04 (1360 days), buy and hold unless noted. Allocator baselines are walk-forward with shrinkage and turnover cost.}
  \label{tab:backtest}
\begin{tabular}{lrrr}
\toprule
Strategy & CAGR & Sharpe & Max DD \\
\midrule
Equal weight & $22.8\%$ & $\mathbf{1.19}$ & $-27.1\%$ \\
Current weight (ours) & $20.7\%$ & $1.14$ & $-26.8\%$ \\
Risk parity & $18.8\%$ & $1.11$ & $-26.4\%$ \\
Minimum variance & $14.1\%$ & $0.94$ & $\mathbf{-21.7\%}$ \\
Benchmark (SPY) & $15.8\%$ & $0.95$ & $-24.5\%$ \\
\midrule
Rebalanced current & $18.8\%$ & $1.09$ & $-28.0\%$ \\
\bottomrule
  \end{tabular}
\end{table}

\begin{figure}[tb]
  \centering
  \includegraphics[width=\columnwidth]{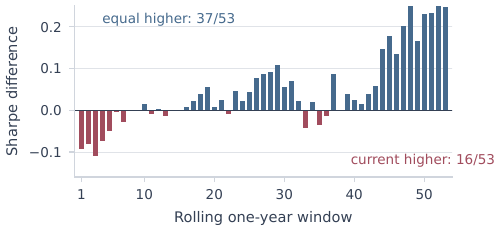}
  \caption{Rolling one-year excess Sharpe often favors equal weighting.}
\Description{Bar chart of the rolling one-year Sharpe difference between equal and current weighting across 53 windows, positive in most windows but with several negative stretches.}
  \label{fig:robustness}
\end{figure}

A single full-sample comparison cannot separate structure from luck, so we probe the equal-versus-current result two ways over the same daily-return matrix, using fixed-weight (daily-rebalanced) portfolios for a clean Sharpe comparison. First, rolling one-year windows stepped monthly: equal weighting has the higher Sharpe in $37$ of $53$ windows ($70\%$; Figure~\ref{fig:robustness}). The negative windows are concentrated in a few stretches rather than spread uniformly. Second, a circular block bootstrap ($2{,}000$ resamples of 21-day blocks, fixed seed) on the full-period Sharpe difference $\mathrm{Sharpe}(\text{equal})-\mathrm{Sharpe}(\text{current})$ gives a point estimate of $+0.053$ with a $95\%$ interval $[-0.024, +0.127]$ and $P(\text{equal better})=91\%$. The direction favors equal weighting, but the magnitude is \emph{not} decisive at the $5\%$ level. On this basket, the tuned weighting does not dominate a simple equal-weight control. That result is the point: the contribution is the audit protocol, and the factor weighting is a replaceable input to it.

\balance
\section{Discussion}

The benchmark makes one claim: whether an action is admissible, grounded, stable, and reproducible before any return is computed. That is the operational boundary that return-based and overlap-based benchmarks lack. Any advisor runs behind the same interface and answers one question: given these dated facts, this portfolio, and this cost model, did it emit a valid, stable, grounded, portfolio-admissible action? The question is small enough to answer exactly and important enough to ask before any performance claim.

The controls establish a precise result: validity, stability, and agreement are separable. A constructed advisor that agrees with the baseline yet violates a constraint is an existence proof that the three axes measure different things, which a single conflated score cannot represent. The bank quantifies that separation at scale, and the model audit shows real systems fail the contract on arithmetic rather than judgment. Separability is the property that an action-level benchmark must report, and it is established by construction rather than estimated as a failure rate.

The agent literature offers a related tension. Live arenas and bias-mitigated long-horizon backtests disagree on whether language-model agents beat passive baselines, but they converge on two points: behavior is governed by the agent's decision and risk logic more than by the model backbone, and the agents' deepest weakness is a missing layer of domain-aware financial logic, not insufficient scale. A deterministic verifier is precisely that layer, made explicit and external. It does not try to out-reason the model; it encodes the admissibility a portfolio actually imposes, so that the unit of evaluation shifts from a model family in the abstract to the action a system emits. The same move is what makes the audit cheap and stable, while return verdicts swing with the horizon and regime.

The empirical calibration sharpens the claim. In fee-quantized markets, diversification has an implementation cost, and the planner should diversify only when the split is fee-neutral. The fundamentals-weighted basket beats the market benchmark but ranks third of five on risk-adjusted return, behind equal weighting and risk parity. That result is consistent with $1/N$ diversification being hard to beat \cite{demiguel2009naive}. The payoff is not a new alpha factor; it is a replayable boundary between generated advice and executable portfolio action.

The protocol carries an explicit interpretation contract. It scores process: advice must be executable, repeatable, grounded in the frozen scenario, and compatible with portfolio constraints before its return means anything. A process metric usually invites Goodhart's law, that optimizing the measure corrupts it, but here, validity, stability, and grounding are the operational requirements themselves, not proxies for them: an advisor that learns never to overrun cash or split sub-tranche orders has, by that fact, learned to act admissibly. The audit certification process is the pre-filter that must be met before outcome evaluation is meaningful. The model audit spans two frontier models. Their bare-prompt failures, arithmetic for the GPT-class model, and abstention for the other, are both largely removed once the fee arithmetic is supplied deterministically. The failure modes it catches also appear in the controls and the bank, where agreement stays high as validity falls. The backtest is a current-basket calibration on free adjusted closes, not a point-in-time selection test, and the cost model covers fixed per-order tranche fees, not spreads, taxes, slippage, or integer lots.

\paragraph{Modular governance contracts.} The contract is modular, and this is a property of the present architecture, not a promise of future work. Write it as a set of decidable predicates $C=\{p_1,\dots,p_n\}$ over a scenario $\sigma$ and a proposed action $a$, with verdict $v=\mathrm{evaluate}(C,\sigma,a)$ that admits $a$ iff every predicate holds. A new governance requirement is one more predicate: an integer-share rule requires $a_i/p_i\in\mathbb{Z}$, a spread budget keeps $s_i a_i$ below the scenario tolerance, a liquidity cap bounds order size as a fraction of frozen volume, a restricted list forbids the scenario's blocked tickers, and per-sector caps join the same set. Tiered commissions and odd-lot surcharges change only the fee function, leaving the no-split rule and the $c/\tau$ floor intact. Such requirements are incorporated as deterministic predicates without changing the audit protocol, the replay mechanism, or the evaluation metrics. The artifact ships one of them, a restricted-list predicate, with a test confirming that an unused clause leaves every existing verdict unchanged. Coverage and governance scale on independent axes: more frozen scenarios widen the bank, and more predicates widen the contract. The defining property still holds, since a recommendation is judged as an executable object before any return is computed. For the AI-finance community, this is a portable target: a shared bank of frozen scenarios and a declarative admissibility contract against which heterogeneous advisors, from prompted LLMs to robo-allocators, are compared on identical, replayable terms, with decision records that double as the audit trail a compliance review already requires.

Pre-trade validation already runs inside robo-advisors and portfolio systems, but as proprietary checks that neither replay nor compare across advisors. The protocol makes that layer shared, declarative, and model-agnostic: identical frozen scenarios and one admissibility contract score any advisor on the same terms, and a rejection is a typed, replayable reason, not an opaque verdict. That is the step from an internal validator to a benchmark the community can audit against.

\section{Conclusion}

The lesson of these experiments outlasts the benchmark: admissibility is an independent property of an AI-generated action, separable from agreement and from return, and on real frontier models, the dominant failure is computation, not judgment, which a deterministic layer removes outright. Auditing a recommendation as an executable object, before any price path is realized, turns a vague sense of trust into a typed, replayable verdict. We release the standard that makes this routine: a deterministic baseline, the audit protocol, and a frozen scenario bank with audited outputs, against which any advisor is judged on identical terms, and into which richer microstructure or compliance rules are incorporated as additional predicates without changing how the audit is scored.

\section*{LLM usage considerations}
The authors used Grammarly and ChatGPT for grammar checking, limited editorial revision, and minor formatting assistance for figures and illustrations. All scientific claims, analyses, experiments, visual representations, and conclusions were produced and verified by the authors.

\bibliographystyle{ACM-Reference-Format}
\bibliography{references}

\end{document}

%% file: figures/audit_boundary.tikz
\begin{tikzpicture}[
  x=1cm,
  y=1cm,
  >=Stealth,
  header/.style={
    font=\sffamily\scriptsize\bfseries,
    text=arenaGray!88!black,
    align=center
  },
  panel/.style={
    rounded corners=3pt,
    draw=arenaInk!20,
    fill=arenaLight,
    line width=.35pt
  },
  card/.style={
    rounded corners=2pt,
    draw=arenaInk!42,
    fill=white,
    line width=.45pt,
    align=center,
    inner xsep=3pt,
    inner ysep=2.4pt,
    font=\sffamily\scriptsize
  },
  bluecard/.style={card, draw=arenaBlue!62, fill=arenaBlue!6},
  greencard/.style={card, draw=arenaGreen!68!black, fill=arenaGreen!6},
  redcard/.style={card, draw=arenaRed!75!black, fill=arenaRed!6},
  rule/.style={
    rounded corners=1.5pt,
    draw=arenaInk!24,
    fill=white,
    line width=.32pt,
    font=\sffamily\tiny,
    align=left,
    inner xsep=3.5pt,
    inner ysep=2.2pt
  },
  wire/.style={draw=arenaInk!70, line width=.5pt},
  arrow/.style={wire, -{Stealth[length=3.4pt,width=4.2pt]}},
  note/.style={font=\sffamily\scriptsize, text=arenaGray!90!black, align=center}
]

% Four groups; symmetric gaps 0.9/0.7/0.9 give equal-length connecting arrows.
\path[use as bounding box] (-.05,-1.52) rectangle (11.60,1.66);

% panels
\node[panel, minimum width=2.05cm, minimum height=2.36cm] at (1.175,0) {};
\node[panel, minimum width=3.40cm, minimum height=2.36cm] (contract) at (6.80,0) {};
\node[panel, minimum width=2.05cm, minimum height=2.36cm] at (10.425,0) {};

\node[header] at (1.175,1.32) {frozen evidence};
\node[header] at (6.80,1.32) {deterministic contract};
\node[header] at (10.425,1.32) {review artifact};

% inputs
\node[bluecard, text width=1.50cm] (scenario) at (1.175,.50)
  {$\sigma$ scenario\\[1pt]{\tiny cash, holdings,\\universe, facts}};
\node[bluecard, text width=1.50cm] (reply) at (1.175,-.50)
  {$r$ reply\\[1pt]{\tiny tickers, dollars,\\citations}};

% parse
\node[greencard, text width=1.30cm] (parse) at (3.75,0)
  {parse $r{\to}a$\\[1pt]{\tiny typed action}};

% contract content
\node[header, text=arenaGreen!50!black] at (6.80,.86)
  {$C(\sigma,a)$ holds iff every row holds};
\node[rule, text width=2.90cm] (admissible) at (6.80,.40)
  {\textbf{Admissible:} universe, cap, cash};
\node[rule, text width=2.90cm, draw=arenaRed!55!black, fill=arenaRed!6] (economic) at (6.80,0)
  {\textbf{Economic:} floor, split, fee};
\node[rule, text width=2.90cm] (grounded) at (6.80,-.40)
  {\textbf{Grounded:} facts, replay hash};
\node[note] at (6.80,-.86) {decided before any return};

% outputs
\node[greencard, text width=1.50cm] (pass) at (10.425,.74)
  {PASS\\[1pt]{\tiny executable orders}};
\node[redcard, text width=1.50cm] (fail) at (10.425,0)
  {FAIL\\[1pt]{\tiny typed violation}};
\node[card, text width=1.50cm] (metrics) at (10.425,-.74)
  {$V,\,S,\,S_w,\,A$\\[1pt]{\tiny audit metrics}};

% input merge: short stub, then equal-length arrow to parse
\draw[wire] (scenario.east) -- (2.51,.50);
\draw[wire] (reply.east) -- (2.51,-.50);
\draw[wire] (2.51,.50) -- (2.51,-.50);
\draw[arrow] (2.51,0) -- (parse.west);

% parse -> contract (same arrow length)
\draw[arrow] (parse.east) -- (contract.west);

% output branch: short stub, then equal-length arrows to PASS / FAIL / metrics
\draw[wire] (contract.east) -- (9.04,0);
\draw[wire] (9.04,.74) -- (9.04,-.74);
\draw[arrow] (9.04,.74) -- (pass.west);
\draw[arrow] (9.04,0) -- (fail.west);
\draw[arrow] (9.04,-.74) -- (metrics.west);
\end{tikzpicture}